\documentclass[twocolumn,english,prb,showpacs]{revtex4-1}
\usepackage{hyperref}
\usepackage[latin9]{inputenc}
\usepackage{amstext,bm}
\usepackage{graphicx}
\usepackage{xcolor}
\usepackage{amsmath}
\usepackage{babel}
\usepackage[titletoc]{appendix}
\usepackage{hyperref}
\usepackage{float}

\begin{document}

\title{{Interaction-driven Quantum Anomalous Hall Effect in Halogenated Hematite Nanosheets}}

\author{Qi-Feng Liang$^{1}$}\email{qfliang@usx.edu.cn} \author{Jian Zhou$^{2}$} \author{Rui Yu$^{3}$} \author{Xi Wang$^{4}$} \author{Hongming Weng$^{5, 6}$}
\email{hmweng@iphy.ac.cn}
\address{$^{1}$ Department of Physics, Shaoxing University, Shaoxing 312000, P. R. China}
\address{$^{2}$ National Laboratory of Solid State Microstructures and Department of Materials Science and Engineering, Nanjing University, Nanjing 210093, China}
\address{$^{3}$ School of Physics and Technology, Wuhan University, Wuhan 430072, P. R. China}
\address{$^{4}$ School of Sciences, Beijing Jiaotong University, Beijing 100044, P. R. China}
\address{$^{5}$ Beijing National Laboratory for Condensed Matter Physics, and Institute of Physics, Chinese Academy of Sciences,Beijing 100190,  P. R. China}
\address{$^{6}$ Collaborative Innovation Center of Quantum Matter, Beijing, China}

\begin{abstract}
Based on first-principle calculations and $k\cdot p$ model analysis, we show that the quantum anomalous Hall (QAH) insulating phase can be realized in the functionalized hematite (or $\alpha$-Fe$_2$O$_3$) nanosheet and the obtained topological gap can be as large as $\sim$300 meV. The driving force of the topological phase is the strong interactions of localized Fe 3$d$ electrons operating on the quadratic band crossing point of the non-interacting band structures. {Such interaction driven QAH insulator is different from the single particle band topology mechanism in experimentally realized QAH insulator, the magnetic ion doped topological insulator film.} Depending on the thickness of the nanosheet, topological insulating state with helical-like or chiral edge states can be realized. Our work provides a realization of the interaction-driven QAH insulating state in a realistic material.
\end{abstract}
\maketitle

\noindent\textit{Introduction.---}
The recently discovered quantum anomalous Hall (QAH) insulating state is a novel topological phase where chiral edge states conduct electric currents dissipationlessly along the sample edge.\cite{Haldane_1988, Yuruiscience, QAHEscience, ChernSM_XuGang, RanyingPRB, LQF_NJP, QiaoQAHE, QAHEvanderbilt,WengAdvInPhys2015} At present, the dissipationless transporting of QAH state in magnetically-doped topological insulator is observed only at extremely low temperature\cite{QAHEscience} and hence its working temperature needs to be raised substantially. According to the seminar work of Thouless \textit{et al.}~\cite{TKNN1982PRL}, the number of the chiral edge states is indexed by the TKNN number or Chern number given by the integrals of the Berry curvature of occupied states in the whole Brillouin zone. In such non-interacting scenario, large spin-orbit coupling (SOC) \cite{Yuruiscience, ChernSM_XuGang, RanyingPRB, LQF_NJP, QiaoQAHE, QAHEvanderbilt} or noncollinear spin textures\cite{ZhouLiang2016PRL} are usually needed to engineer the nontrivial Berry phase in the realistic material systems. In parallel with the QAH insulators within non-interacting framework, there have been growing interests in exploring it and other topological states in interacting systems,\cite{Raghu2008PRL,SunKai_PRL,Zhu2016Interaction, Wu2016Diagnosis,XiDai2014PRB,Murray2014Renormalization,BL111PRL} where the electron interactions become the driving force. If the interaction-driven QAH state (IQAH) is realized, it will not only significantly expand the possible material systems of QAH phase, but also apparently deepen our understanding about the topologically nontrivial phases of many-body systems.

It has been proven that the electron interactions can induce nontrivial topology in the ground state wavefunction of an interacting system. Raghu and co-workers \cite{Raghu2008PRL} found that in a half-filling honeycomb lattice, the next-nearest-neighbor repulsive interaction drives the system into a QAH insulating state. Mean-field model studies show that the resultant QAH state is favored by the Dirac fermions only when the second-nearest-neighbor interaction is the leading repulsive force and larger than some critical value, a condition which is usually not satisfied in realistic material. For instance, in a possible candidate material of the interaction-driven Dirac-QAH state\cite{Xiao2011Interface,LQF_NJP,XiDai2014PRB}, the bilayer heterostructure of transition metal oxides, experimental investigation did not observe the expected QAH insulating state but a trivial Mott anitferromagnetic (AFM) ground state\cite{BL111PRL}. Detailed numerical studies found that the quantum fluctuations strongly suppress the QAH order and the ground state is more likely to be driven into other competing orders\cite{MotrukPRB}. To overcome the limitations of Dirac fermions, Sun \textit{et al.} \cite{SunKai_PRL} proposed an interacting model which starts from a quadratic band crossing point (QBCP) rather than the Dirac point. In contrast to the Dirac point, the QBCP in two-dimension has a constant density of states (DOS) at the Fermi level, and also carries a nontrivial 2$\pi$ Berry phase. This feature makes the QBCP fermionic system marginally unstable and easily be driven into the QAH state even by arbitrarily weak interaction.\cite{XiDai2014PRB,RueggPRB} Such encouraging improvement of the IQAH state has recently been confirmed by the unbiased numerical exact diagonalization studies.~\cite{Zhu2016Interaction, Wu2016Diagnosis} Despite of these theoretical progress, there still lacks material realization of this interaction driven topological state.

Here we predict that the recently synthesized hematite (HM) nanosheets,\cite{WXHMNS} if functionalized by halogen atoms or hydroxyl (OH) groups, would realize a QAH state with topological gap as large as $\sim$300 meV. Our hybrid density functional calculation shows that electron interaction is the driving force of the QAH state. {The band structure within the non-interacting single-particle approximation contains a QBCP at the $\Gamma$ point composed of the $d_{xz}$ and $d_{yz}$ orbitals of Fe atoms. A $k\cdot p$ model based on these orbitals describes well the low-energy behaviors around the QBCP.} Turning on the electron interaction leads to a logarithmical divergence in the susceptibility of the QAH order, implying a strong tendency to the QAH insulating state. Therefore, our system is indeed a material candidate for the IQAH proposed in the QBCP fermionic systems.\cite{SunKai_PRL} Benefited from the ultra-high N$\acute{\text{e}}$el temperature of hematite ($T_{N}\sim$1000 K) and the large topological gap, the predicted IQAH is expected to work safely above the room temperature.

\noindent\textit{Methods.---}
The first-principle calculations were done with the Vienna \emph{ab initio} simulation package (VASP) \cite{VASP}, where projected augmented-wave (PAW) potential is adopted. \cite{PAW_Blochl,PAW_Kresse_1999} We apply the exchange-correlation functional introduced by Perdew, Burke, and Ernzerhof (PBE)~\cite{PBE} within generalized gradient approximation (GGA) in the calculations. For better treatment of the electron interaction, the hybrid density functional (HSE06) is alternatively adopted \cite{HSE06}. In all the calculations the energy cutoff for the plane-wave basis is set as 520 eV and the forces are relaxed less than 0.01 eV/$\rm \text{\AA}$. The tight-binding model of the nanoribbon structure is constructed by using the Maximally Localized Wannier Functions (MLWF) method coded in WANNIER90.\cite{Wannier90}

\begin{figure}
\centering{}
\includegraphics[width=0.45\textwidth]{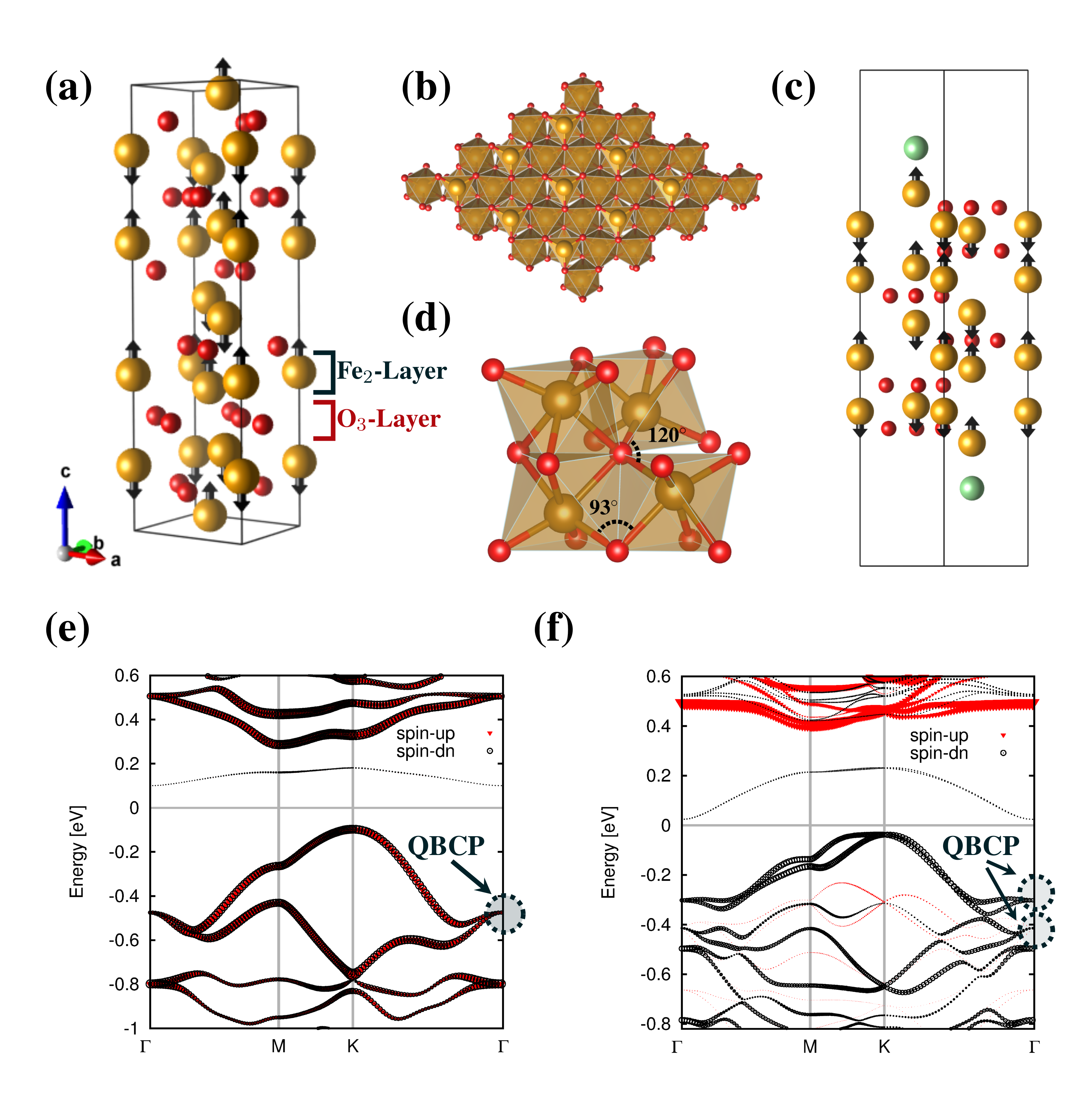}
\caption{ (Color online)
(a) The hexagonal unit cell of $\alpha-$Fe$_2$O$_3$. The Fe and O atoms are denoted by brown and red balls. The ``up-up-down-down" AFM arrangement of Fe ions are denoted by arrows.(b) Top view of the exposed \{0001\} facet of Fe$_2$O$_3$. (c) Side view of a X/HM$_5$/X nanosheet with X being halogen atoms or OH groups. The X atoms are denoted by cyan balls at the on-top site of outmost Fe atoms. (d) Fe-O-Fe bond angles within and between the Fe bilayers. (e) and (f) show the GGA band structures of HM$_2$ and HM$_5$ nanosheets, respectively. {The weight of 3$d$ orbitals on second outmost Fe atoms (see also in Fig.\ref{fig:order2}(a)) in each eigenstate is shown by the thickness of the curves.} The QBCPs at the $\Gamma$ point are high-lighted by dotted circles.
\label{fig:crystal}}
\end{figure}
\vspace{3mm}
\noindent\textit{Hematite Nanosheets.---}
The hematite, or $\alpha$-Fe$_2$O$_3$, is one of the most abundant natural minerals in the earth's crust. Its crystal adopts a corundum structure with the space group of $R\bar{3}c$. According to the positions of atoms, the HM crystal can be considered as an alternative stacking of buckled Fe$_2$ bilayers and planar O$_3$ layers along the [0001] direction (see in Fig.\ref{fig:crystal}(a)). Below the N$\acute{e}$el temperature ($T_N=953$ K), the HM crystal has an ``up-up-down-down" antiferromagnetic ground state, in which the high-spin Fe$^{3+}$ ions couple ferromagnetically within the Fe bilayer and antiferromagnetically to the neighboring Fe bilayer (see Fig.\ref{fig:crystal}(a) and (d)). A Goodenough-Kanamori superexchange mechanism~\cite{goodenough1963,kanamori1959} explains the origin of the AFM state very well: All the Fe-O-Fe bond angles within the Fe bilayer are $\sim$93$^\circ$ (see in Fig. \ref{fig:crystal}(d)) and hence the Fe local moments favor the ferromagnetic coupling, whereas between the Fe bilayers the Fe-O-Fe bond angles take larger values such as 120$^\circ$ and 130$^\circ$, which favor the AFM coupling. If the bulk crystal is cleaved, nanosheets with Fe-terminated surfaces are obtained, which can be formally written as the Fe-O$_3$-[Fe$_2$-O$_3$]$_n$-Fe (HM$_n$) nanosheets. Our first-principle calculation also verifies that the bulk ``up-up-down-down" AFM magnetic structure has been inherited to the nanosheets system (see Supplement Informations (SI)). Depending on the nanosheet thickness, there exists two distinct AFM magnetic orders regarding their magnetizations at the surfaces, \textit{i.e.} the parallel magnetizations at the top and bottom surfaces for even HM$_{2n}$ nanosheet and the antiparallel magnetization for odd HM$_{2n+1}$ nanosheet. {In both cases, the band structure within GGA (see in Fig.\ref{fig:crystal}(e) and (f)) shows them an AFM insulator with QBCPs at $\Gamma$ just below the Fermi level. Such QBCPs spanned by $d_{xz}$ and $d_{yz}$ orbitals of Fe are protected by C$_3$ rotation symmetry \cite{Liang2016Top} together with the composite symmetry composed by the spin $\pi$-rotation about the $\hat{s}_x$ axis and the time-reversal symmetry. {Each surface of the HM$_n$ nanosheet contains one QBCP. The spin of the QBCP is determined from the magnetization on its locating surface. For even HM$_{2n}$ nanosheet the two QBCPs are spin-degenerate, while for odd nanosheet they are of the same spin.} To bring the QBCP to the Fermi level, we propose to remove one electron from the system by adsorbing one halogen atom, or [OH]$^{-1}$ group to each surface. The functionalized HM nanosheet can therefore be denoted as X/HMn/X with X being the attached atom on its top and bottom surfaces. Thus, within noninteracting single-particle approximation, one can achieve the semimetal state with QBCP at $\Gamma$.}


\begin{figure}
\centering{}
\includegraphics[width=0.45\textwidth]{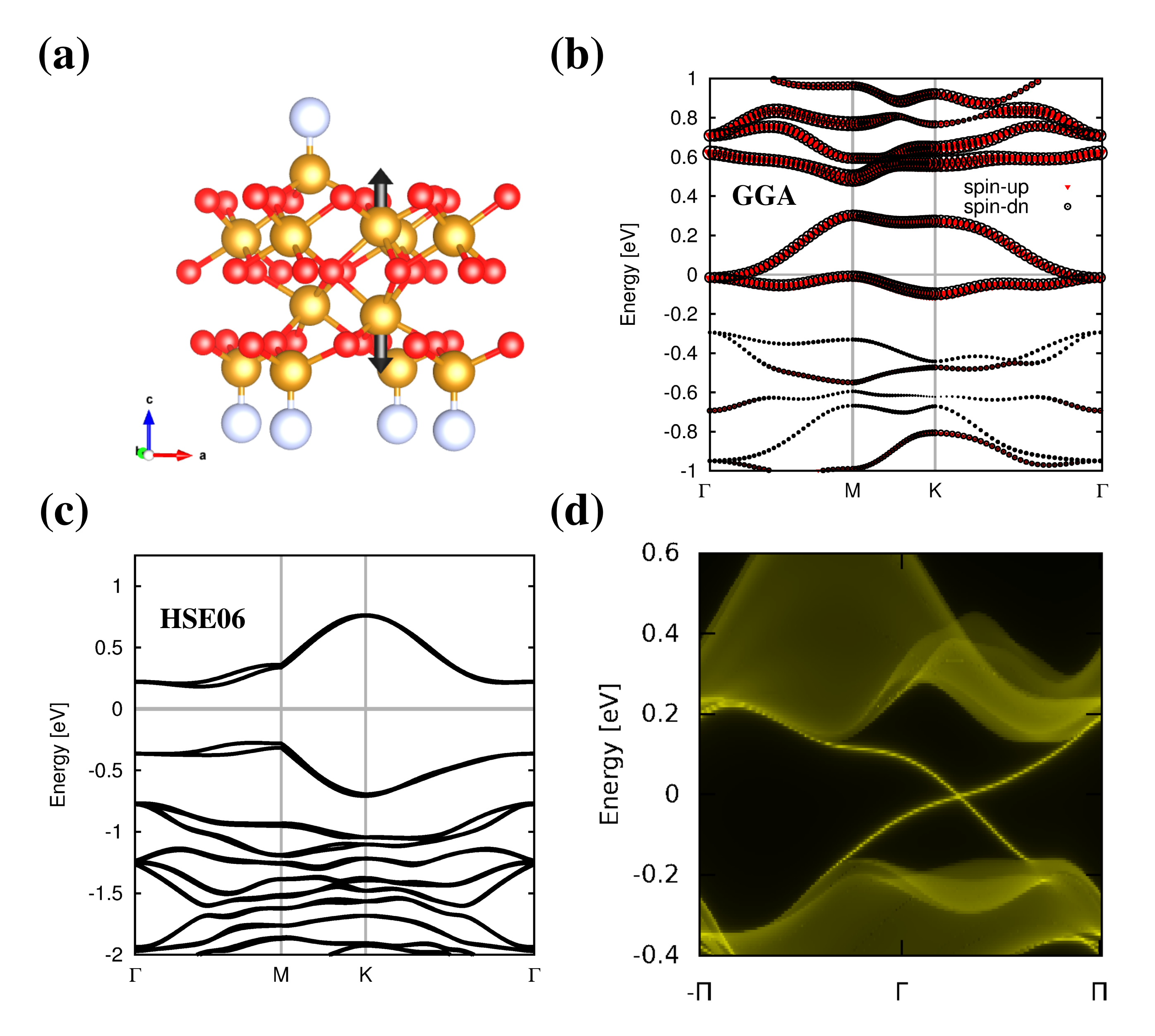}
\caption{(Color online)
{(a) Crystal structure of the F/HM$_{2}$/F nanosheet. Arrows represent the local magnetic momentum on the two second outer-most Fe atoms which dominate the electronic state around the QBCP. (b) GGA band structure without SOC. (c) Band structure with the HSE06 functional. (d) MLWF based tight-binding band structure a F/HM$_{2}$/F ribbon.}
\label{fig:huc-qahe}}
\end{figure}
\vspace{3mm}
\noindent\textit{IQAH states in HM Nanosheets.---}
Here we show that when fully functionalized by halogen atoms, as shown in Fig.\ref{fig:huc-qahe}(a), the {X/HM$_n$/X} nanosheet will be driven into a QAH insulating state by the electron-electron interactions. For simplicity, we only show the result of the F/HM$_n$/F nanosheet and other X/HM$_{n}$/X nanosheets (X = Cl and OH group)\footnote{For X=Br and I, the QBCP is contributed mainly from the $p_x$ and $p_y$ orbitals of Br or I atoms. A QAH gap is opened at the QBCP by the strong SOC of Br or I atoms but not the electron interaction of Fe. The mechanism of QAH in Br/HM/Br and I/HM/I is thus a traditional non-interaction one and completely different from the interaction-driven mechanism in the present work. We may introduce this physics in another work} {have the similar results}. The GGA band structure without SOC is plotted in Fig.\ref{fig:huc-qahe}(b) where one can see F/HM$_2$/F nanosheet is a zero-gap {semimetal} with the conduction and valence bands touching quadratically at the $\Gamma$ point. The Fermi energy passes right through this QBCP. Atomic orbital projection of the electronic states near the QBCP indicates that they are mainly composed of the $d_{xz}$ and $d_{yz}$ orbitals of the second outer-most Fe atoms as marked by arrows in Fig.\ref{fig:huc-qahe}(a). The degeneracy of the $d_{xz}$ and $d_{yz}$ orbitals at the $\Gamma$ point is protected by the  $C_3$ rotation symmetry of the nanosheets\cite{Liang2016Top} without the SOC. One also finds the bands are all spin degenerate because of the nanosheet is now an ideal AFM insulator and the two copies of QBCPs are located on different surfaces. Inclusion of the SOC only develops a negligible gap of 2 meV at the QBCP.

We further adopt the Heyd-Scuseria-Ernzerhof (HSE06) hybrid density functional \cite{HSE06} to treat the electron exchange-correlation interaction more accurately beyond the plain GGA theory. The corresponding band structure is shown in Fig.\ref{fig:huc-qahe} (c), from which one finds the degeneracy at $\Gamma$ is lifted and a large gap of 300 meV is opened. As hinted by previous model studies on interacting QBCP fermions\cite{SunKai_PRL,Zhu2016Interaction, Wu2016Diagnosis}, the gap opened by the hybrid density functional should be topologically nontrivial. This fact is readily verified by calculations of a ribbon sample with the tight-binding model constructed from the MLWF method. In Fig.\ref{fig:huc-qahe}(d) we plotted the energy spectrum of a semi-infinite nanosheet obtained from a recursive Green's function calculation with one edge. One can see a bulk gap of 300 meV and inside of it there exist two helical-like conducting edge states connecting the conduction and valence bands. {These two edge states are from the edges on the top and bottom surfaces, respectively. They have opposite chirality since the opposite magnetic moment orientation as shown in Fig.\ref{fig:huc-qahe}(a).} A possible usage of these helical-like edge states is that they allow the spin-half superconducting pairing and thus can host the Majorana fermions, a basic element of topological quantum computation.\cite{AliceaMF2012} One may worry about that the helical-like edge states in F/HM$_{2n}$/F nanosheets are more vulnerable than that of the Z$_2$ QSH insulators since the time reversal symmetry is broken here. We argue that such helical-like edge states actually could be more stable as long as the nanosheet is thick enough to block out the spin-flip back-scattering between the chiral edge states on top and bottom surfaces.

\begin{figure}
\centering{}
\includegraphics[width=0.45\textwidth]{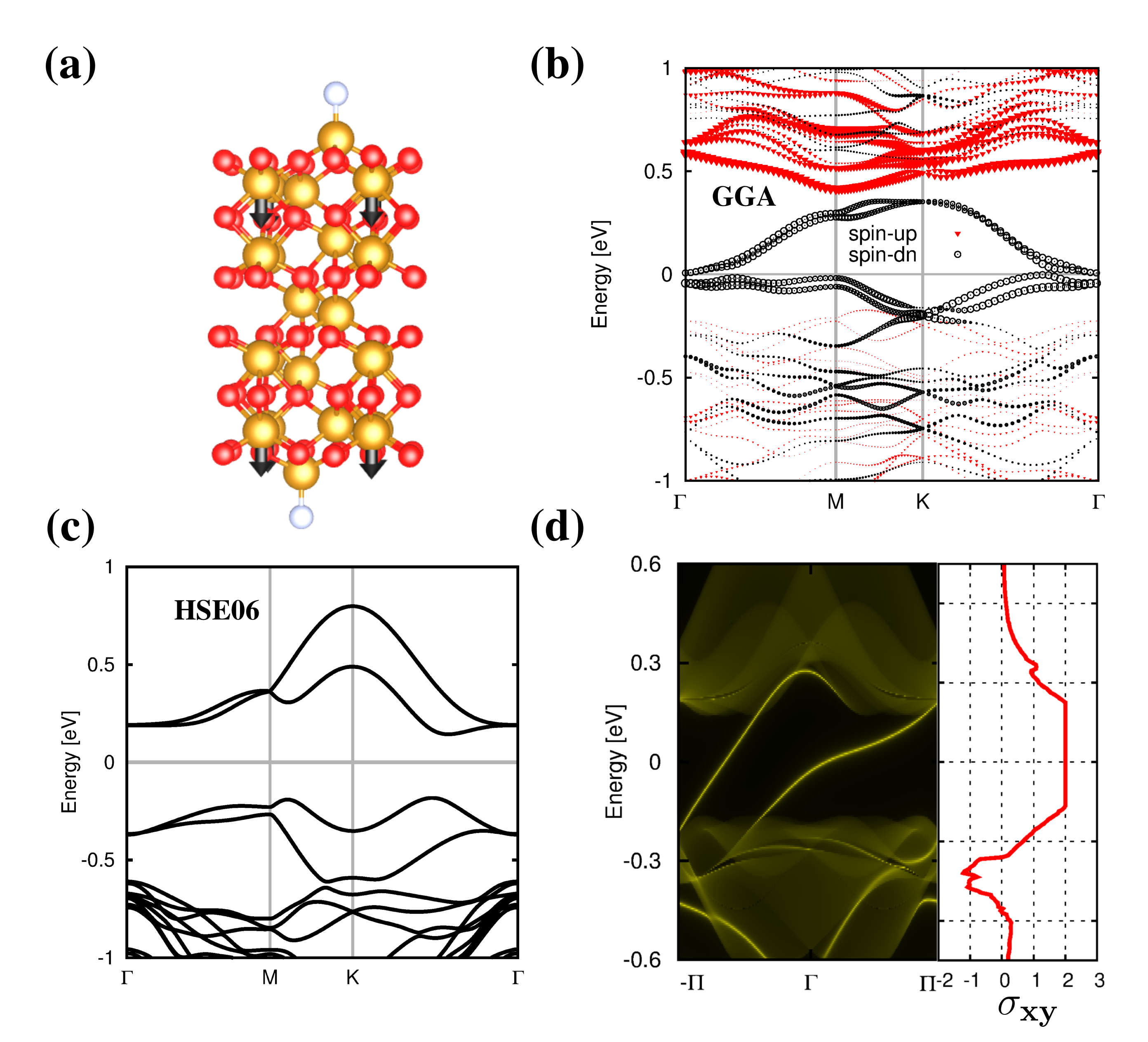}
\caption{(Color online)
{(a) Crystal structure of F/HM$_5$/F nanosheet. The ball (white) on the top-site of Fe denote the F atoms. Arrows mark the local magnetic moment on the Fe atoms which dominate the electronic state of the QBCPs. (b) GGA band structure of the F/HM$_5$/F nanosheet without SOC. The weight of the $d$-orbitals from the Fe atoms marked by arrows in (a) is represented by the thickness of the curves. (c) Band structure with HSE06 functional. (d) Tight-binding band structure of a F/HM$_5$/F ribbon cutting along the $\textbf{a}$ direction. Right pannel shows the calculated anomalous Hall conductance of the ribbon in the unit of $e^2/h$.}
\label{fig:uc}}
\end{figure}

In contrast, in the odd number F/HM$_{2n+1}$/F nanosheets the $d_{xz}$ and $d_{yz}$ orbitals of the Fe atoms on top and bottom surfaces generate two copies of QBCPs with the same spin polarization (See Fig.\ref{fig:uc}(b)) and hybridization betwen them is allowed though it can be neglected if the sheet is quite thick. Within the hybrid density functional approximation the two QBCPs both again open a gap of 300 meV and their Chern numbers are added up to a value $\mathcal{C}=2$. Therefore, two chiral edge states would flow in the same direction around the sample's boundary. It is found that inside the energy window of the topological bulk gap, the Hall conductance is quantized into two unit of ${e^2}/{h}$ as shown in the right of Fig.\ref{fig:uc}(d), confirming the Chern number $\mathcal{C}=2$ of the system.

The above analysis lead to a conclusion that the F/HM$_n$/F nanosheet is indeed driven into the QAH insulating state when the electron interactions are treated with the hybrid density functional. In previous model studies one found the interaction-driven QAH effect is obtained when one treats the electron interactions in the Hartree-Fock approximation. In the hybrid density functional scheme, a fraction of the short-range PBE exchange energy is mixed by the exact exchange of the Hartree-Fock theory\cite{Kummel08RMP}. This explains why only when the hybrid density functional is introduced instead of the plain GGA method, does the QAH effect take place. The mixture fraction of the exact exchange $\alpha=0.25$ (HSE06 version) used in the present work is evidently a practical choice for most materials. We have also checked that even for a relatively smaller mixture $\alpha=0.1$, a topological gap larger than 100 meV can be obtained, implying the robustness of the QAH phase against the choice of $\alpha$.

\begin{figure}[t]
\centering{}
\includegraphics[width=0.46\textwidth]{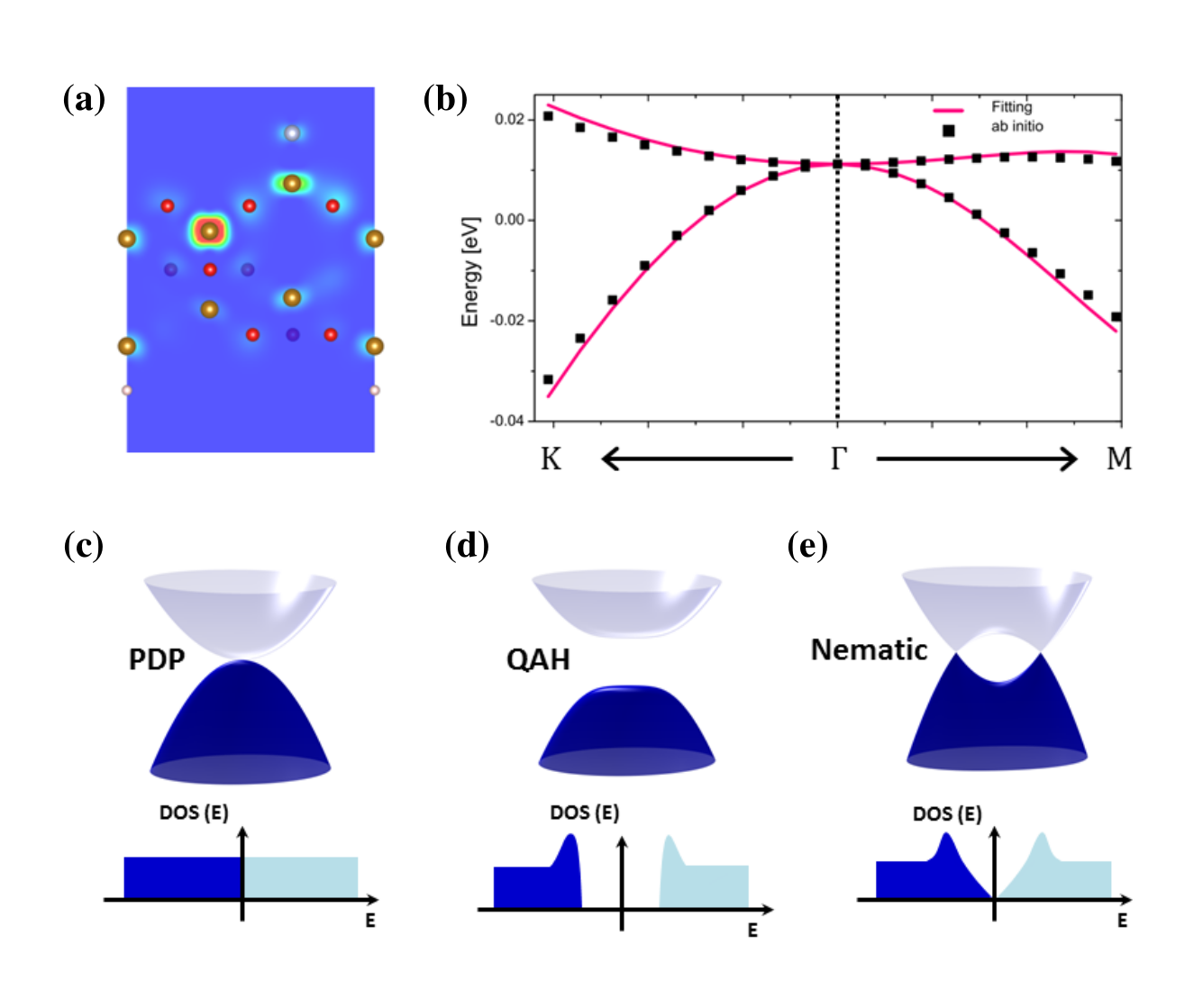}
\caption{(Color online)
(a) Charge density of the electronic state from the QBCP. (b) Fitting of the \textit{ab initio} band structure by the effective $k\cdot p$ model.
(c) Schematic gapless band structure near the QBCP without any $O_i$ ($i$= $x$, $y$, $z$) order parameter. The DOS near the Fermi level is a constant. (d) The gapped band structure and the DOS near $\Gamma$ with the QAH order.  (d) Band structure and the DOS near $\Gamma$ with the nematic order ($O_x$ or $O_y$ order parameter). In this case, QBCP is split into two Weyl points which give rise to a linear DOS proportional to energy E at the low energy regime,
\textit{i.e.} DOS($E$)$\propto E$.
\label{fig:order2}}
\end{figure}

\vspace{3mm}
\noindent\textit{The k$\cdot$p model.---}
To obtain a detailed insight into the mechanism of such large topology gap, we construct an effective two-orbital model based on the $d_{xz}$ and $d_{yz}$ orbitals of the Fe atoms (see the SI), and perform a mean-field level analysis on the mechanism underlying the IQAH state. Up to the forth order of momentum $k$, the non-interacting Hamiltonian without SOC reads,
  \begin{eqnarray}
  H_0(\bm{k})=\left[\begin{array}{cc}
                ak^2+bk^4, & (\gamma+\gamma_1k^2) k_{-}^2+\gamma_2 k_{+}^4\\
                \dag, & ak^2+bk^4
              \end{array}\right]+O(k^6),\nonumber\\
              \label{eq:eff_model}
  \end{eqnarray} where $k_{\pm}=k_x\pm i k_y$. The spin indices are omitted because of the half-metal nature of the QBCP on one surface. As shown in Fig.\ref{fig:order2}(b), the \textit{ab initio} band dispersion around the $\Gamma$ point is fitted very well by the effective Hamiltonian of Eq.(\ref{eq:eff_model}) with suitable choices of $a$, $b$, and $\gamma_i$.

Now we turn on the electron interactions. We only consider the onsite interactions since usually they are much larger than the repulsive interactions with longer range. Thus, the only permitted interaction is $H_{I}$=$V\hat{\psi}_{+}^\dag\hat{\psi}_{+}\hat{\psi}_{-}^\dag\hat{\psi}_{-}$ with $ \hat{\psi}^\dag_{\pm}$ and $\hat{\psi}_{\pm}$ denoting the creation and annihilation operators for the basis states $|\pm\rangle$. Originally this term stems from the repulsive force between the $d_{xz}$ and $d_{yz}$ orbitals on the same Fe site. By applying the Hartree-Fock approximation, the total Hamiltonian $H=H_0+H_I$ is readily reduced to,
\begin{eqnarray}
  H_{MF}=H_0-\frac{V}{2}\left[\begin{array}{cc}
               O_{z}, & O_x+i O_y\\
                \dag, & -O_{z}
              \end{array}\right],
              \label{eq:mf_model}
\end{eqnarray}
where the order parameters $O_i=\langle\hat{\Psi}^{\dag}\sigma_i\hat{\Psi}\rangle$ ($i$ =$x$,$y$,$z$) with $\sigma_i$ being the Pauli matrices and $\hat{\Psi}=\left[\begin{array}{c}\hat{\psi}_{+}\\\hat{\psi}_{-}\end{array}\right]$.
The $O_z$ order parameter is found to give rise to a band gap at the QBCP (see Fig.\ref{fig:order2}(d)) and the resulted gapped state has a nonzero Chern number $\mathcal{C}=1$. In the cases of $O_x$ or $O_y$, which are also termed as nematic orders, the QBCP is split into two Weyl points and the system remains gapless as shown in Fig.\ref{fig:order2}(e) schematically. The tendencies to these orders are determined by their susceptibilities defined as $\chi_i=\frac{\partial^2 E_{tot}}{\partial O_i^2}$, where the total energy $E_{tot}$ is given by the summation of the energies of all occupied states. At the weak-coupling limit, one easily verifies that all three susceptibilities diverge logarithmically, indicating these symmetry-breaking orders would emerge even at infinitsemal interactions. One further verifies there exists a relation $\chi_z=\chi_x+\chi_y$, ensuring that the QAH order is the leading instability as found by our $\textit{ab initio}$ calculations. The reason why the QAH order dominates over the nematic orders is understood from the different DOS profiles of the system within these orders. When QAH order is developed, a full energy gap is opened and the system gains kinetic energy from this gap opening (see the bottom of Fig.\ref{fig:order2}(d)). In contrast, when nematic orders take place the QBCP is split into Weyl points, which only deforms the DOS to a "V"-shaped profile (see the bottom of Fig.\ref{fig:order2}(e)) rather than a full gap, making less gain of energy.


\vspace{3mm}
\noindent\textit{Discussion.---}
The interaction QBCP fermions can also be created in the bilayer heterostructure of transition metal oxides\cite{RueggPRB,XiDai2014PRB}. It is found that the IQAH state is possible only when the interaction strengths and spin states are suitably chosen and the aimed magnetic ground state is achieved. Also the fabrication of high quality bilayer heterostructure of the transition metal oxides along the polar [111] direction is still a challenging task. For the present system, the bulk AFM ground state with ultra-high N$\acute{\text{e}}$el temperature is well verified and this AFM ground state is expected to be maintained in the nanosheet system as supported by our first-principles calculations. The QAH state found is very stable irrelevant to the choice of  mixture parameter in HSE scheme, the strength of SOC and the direction of spin axis. Also the HM$_{m}$ nanosheet is easily synthesized through a cheap wet-chemistry method\cite{WXHMNS}. Moreover, the present system takes the advantages of quite large topological gap, rich in the earth's crust, providing a fascinating platform for the IQAH states.

\vspace{3mm}
\noindent\textit{Acknowledgements.---}
We thank the valuable discussion with Ziyang Meng and Chen Fang. The work is supported by National Natural Foundation of China (NFSC) (Grants No.11574215, No. 11274359 and No. 11422428). Q.F.L acknowledges the support from the Scientific Research Foundation for the Returned Overseas Chinese Scholars, State Education Ministry of China. H.M.W is supported by the National Key Research and Development Program of China (Grant No. 2016YFA0300600), and the ``Strategic Priority Research Program (B)" of the Chinese Academy of Sciences (Grant No. XDB07020100). The calculations in this work were performed on the supercomputer of the Shanghai Supercomputer Center and the computer facilities in the High Performance Computing Center of Nanjing University. Q.F.L and J.Z contributed equally to this work.

\bibliographystyle{apsrev4-1}
\bibliography{refs}

\end{document}